\documentclass[a4paper]{article}

\usepackage{INTERSPEECH2019}

\ninept

\usepackage{graphicx}
\usepackage{amsmath,amssymb}

\usepackage{anyfontsize}
\usepackage{subfig}

\newcommand{\Lcal}{\mathcal{L}}
\def\E{{\mathbb{E}}}
\newcommand{\norm}[1]{\left\lVert#1\right\rVert}

\usepackage[]{hyperref}
\hypersetup{
    colorlinks = false,
    hidelinks = true,
}

\title{GELP: GAN-Excited Linear Prediction for Speech Synthesis from Mel-spectrogram
}
\name{Lauri Juvela$^1$,  Bajibabu Bollepalli$^1$, Junichi Yamagishi$^{2,3}$, Paavo Alku$^1$}
\address{
  $^1$Aalto University, Finland \\
  $^2$National Institute of Informatics, Japan; 
  $^3$University of Edinburgh, UK}
\email{\{lauri.juvela, bajibabu.bollepalli, paavo.alku\}@aalto.fi, jyamagis@nii.ac.jp}

\begin{document}

\maketitle
\begin{abstract}
 Recent advances in neural network -based text-to-speech have reached human level naturalness in synthetic speech. The present sequence-to-sequence models can directly map text to mel-spectrogram acoustic features, which are convenient for modeling, but present additional challenges for vocoding (i.e., waveform generation from the acoustic features). High-quality synthesis can be achieved with neural vocoders, such as WaveNet, but such autoregressive models suffer from slow sequential inference. Meanwhile, their existing parallel inference counterparts are difficult to train and require increasingly large model sizes. In this paper, we propose an alternative training strategy for a parallel neural vocoder utilizing generative adversarial networks, and integrate a linear predictive synthesis filter into the model. Results show that the proposed model achieves significant improvement in inference speed, while outperforming a WaveNet in copy-synthesis quality. 
\end{abstract}
\noindent\textbf{Index Terms}: Neural vocoder, Source-filter model, GAN, WaveNet 
\section{Introduction}

In recent years, the state of the art in text-to-speech (TTS) synthesis has been advancing rapidly. Introduction of deep neural networks for acoustic modeling \cite{Zen2013-DNN} in statistical parametric speech synthesis (SPSS) \cite{Zen2009-SPSS} started the ongoing trend of deep learning in TTS, and was soon followed by recurrent networks \cite{fan2014tts}, and more recently, sequence-to-sequence models with attention \cite{Sotelo2017char2wav, Wang2017-tacotron}. The latter learn to align input text (or phoneme) sequences with the target acoustic feature sequences, which in effect integrates a duration and acoustic model into a single neural network.
Meanwhile, the advances in neural network models for speech waveform generation, such as WaveNet \cite{oord2016-wavenet}, have raised the bar in synthetic speech quality. Although the original WaveNet was controlled mainly with linguistic features, it used supplementary acoustic information from  duration and pitch predictor models. The role of a waveform generator model has since shifted more to a ``neural vocoder'' \cite{Tamamori2017-wavenet-vocoder}, i.e., the model is conditioned directly on acoustic features. With this division of labor, neural vocoders have been shown to produce high-quality synthetic speech when paired with a sequence-to-sequence model \cite{Shen2018-tacotron2}, or a more conventional SPSS pipeline \cite{Wang2018-comparison-of-waveform-generation}.

However, WaveNet and similar models suffer from their sample-by-sample sequential inference process. Although the inference can be made faster with dilation buffering \cite{Ramachandran2017-fast-generation-cnn}, or optimizing recurrent neural networks (RNNs) for real-time use \cite{kalbrenner2018-wave-rnn}, sequential inference remains ill-suited for massively parallel applications. To remedy this, models capable of parallel inference have been proposed \cite{oord2017-parallel-wavenet, ping2018-clarinet-parallel-wave-generation}. The training setup involves two models: a pre-trained autoregressive model (teacher) is used to score the outputs of a feed-forward parallel model (student). The student training criterion is related to inverse autoregressive flows, and alternative flow-based models have been further presented \cite{Prenger2018-waveglow}. A major issue with these parallel synthesis models is their restriction to invertible transforms, which limits model capacity utilization. This in turn leads to large models that require a considerable amount of training \cite{ping2018-clarinet-parallel-wave-generation, Prenger2018-waveglow}.

%
%
In addition to flow-related scoring rules, more conventional short time Fourier transform (STFT)-based regression losses have been found essential in training parallel waveform generators \cite{oord2017-parallel-wavenet, ping2018-clarinet-parallel-wave-generation}. Furthermore, similar models have  been trained solely using STFT-based losses \cite{wang2019-neural-source-filter-model}. The method builds on the classical source-filter model of speech, and learns a neural filter for a mixed excitation signal constructed from harmonic and noise components \cite{wang2019-neural-source-filter-model}. 
However, this approach requires explicit voicing and pitch inputs, and is sensitive to the alignment between generated and target signals when STFT phase loss is used.
%
%
Meanwhile, the source-filter model has also been used with linear predictive (LP) filters to build neural models for the glottal excitation  \cite{juvela2018-glotnet-interspeech, juvela2019-glotnet-taslp} or the LP residual signal  \cite{valin2019-LPCNet, Hwang2018-LP-wavenet}.
The use of LP inverse filtering can be seen analogous to the the flow-based approaches, as they apply an invertible transform whose residual is more Gaussian and is thus easier to model \cite{juvela2019-glotnet-taslp}. However, these LP-based models still use sequential inference (similar to WaveNet or WaveRNN) and the inherent speed limitations remain in place. 

Generative adversarial networks (GANs) \cite{goodfellow2014generative} are especially appealing for waveform synthesis, 
as they enable unrestricted use of feedforward architectures capable of parallel inference. 
%
GAN-based residual excitation models have been proposed previously for all-pole envelopes derived from mel-frequency cepstral coefficients (MFCCs)  \cite{juvela2018-synthesis-from-mfcc}, with further refinements to the GAN architectures and training procedure presented in \cite{juvela2019-multiscale-gan-synth}. However, these models operate in a pitch-synchronous synthesis framework, which makes them sensitive to pitch marking accuracy during training, and requires pitch information at synthesis time. Further, the models were optimized solely in the excitation domain, which may be perceptually sub-optimal \cite{valin2019-LPCNet, Hwang2018-LP-wavenet}.

In the present Tacotron style TTS systems \cite{Shen2018-tacotron2}, the most commonly used acoustic feature is the log-compressed mel-spectrogram, that utilizes the classical MFCC triangular filterbank \cite{davis1980mfcc}. The mel-spectrogram is appealing due to its simple estimation, perceptual relevance, and overall ubiquity in speech applications. However, from a TTS perspective, the mel-spectrogram poses additional challenges due to its lack of explicit voicing and pitch information. Nevertheless, the typically used 80 mel filters have enough frequency resolution to include some harmonics, which is sufficient for rudimentary waveform reconstruction  \cite{boucheron2012low-bitrate-mfcc-codec}.
%
Altogether, there remains a  need for a fast high-quality neural vocoder capable of generating speech waveforms directly from mel-spectra with relatively little training. 

In this paper, we combine a MFCC-based envelope model \cite{juvela2018-synthesis-from-mfcc} with recent GAN training insights \cite{juvela2019-multiscale-gan-synth}. Furthermore, we use neural net architectures that operate directly on raw audio, and integrate a parallel inference capable LP synthesis filter into the computation graph. Inspired by classic speech coding, we call the proposed method ``GAN-excited linear prediction'' (GELP). 


\section{Methods}

\begin{figure}
    \centering
    \includegraphics[width=0.9\linewidth, trim={3cm, 15cm, 8cm, 1cm}, clip]{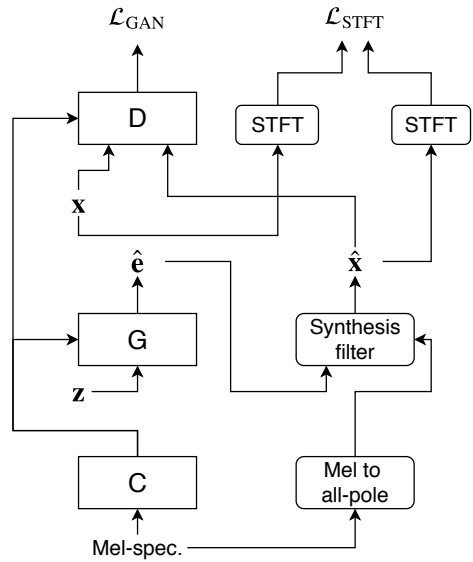}
    \caption{Model training setup. An input mel-spectrogram is passed to a conditioning model $C$, upsampled, and used to control an excitation generator $G$. The generator transforms a white noise input into an excitation signal, which is then filtered with an all-pole spectral envelope extracted from the mel-spectrum. The resulting signal is trained to match a target speech signal by regression on the STFT magnitude and a time-domain adversarial loss provided by discriminator $D$. }
    \label{fig:model_overview}
\end{figure}

Figure~\ref{fig:model_overview} shows an overview of the training setup. The model consists of three trainable components: a generator $G$ and discriminator $D$, both operating at an audio rate, and a conditioning model $C$, operating at a control frame rate. All the models are non-causal 1-D convolution nets, as detailed in section \ref{sec:network_architectures}.
The conditioning model creates a context embedding ($\bm c$)
of the mel-spectrogram at the frame rate, which is linearly upsampled to the audio rate before inputting it into $G$ and $D$. 
The input ($\bm z$) of the generator model is a white noise sequence, sampled at the audio rate, which the model transforms into a LP residual signal $\hat{\bm e}$. 
To produce synthetic speech, the generated residual $\hat{\bm e}$ is fed as an excitation to an LP synthesis filter, whose coefficients are derived from the mel-spectrogram. 

No direct voicing or pitch features are provided to the model, so the model has to infer this information from the input mel-spectrogram.  
%
In contrast, we explicitly use the spectral envelope information contained in the mel-spectrogram. This is achieved by first fitting an all-pole envelope to the mel-spectrum (section \ref{sec:mfcc_inversion}), and second, implementing a synthesis filter suitable for parallel inference (section \ref{sec:stft_synth_filter}).
To train $G$ and $C$, two kinds of loss functions are used: 1) a regression loss based on STFT magnitudes, and 2) a time domain adversarial loss propagated through the discriminator model $D$. The former guides the generator to learn the mel band energies of speech, which is vital for creating harmonics of voiced sounds. The latter loss guides the system to learn in time domain the speech signal's phase information and fine stochastic details.

\subsection{Envelope recovery from mel-spectrogram}
\label{sec:mfcc_inversion}

Spectral envelope fitting to MFCCs has been proposed previously for neural vocoding in \cite{juvela2018-synthesis-from-mfcc}. 
The previous study focused on lower order MFCCs, while a typical Tacotron configuration uses high-resolution mel-spectra with 80 mel filters and omits the discrete cosine transform to cepstrum domain. 
Nevertheless, the method for fitting an all-pole envelope to mel-spectrum is directly applicable. 
The mel-spectrogram $\bm m$ is computed as $\bm m = \log (\mathbf{M} \bm X)$,
%
where $\bm X$  is a pre-emphasized STFT magnitude spectrogram, and $\mathbf{M}$ is a mel-filterbank matrix.
A reconstruction is obtained simply by using the pseudo-inverse of $\mathbf{M}$ and flooring the result with a small positive $\epsilon$ to prevent negative values:
$\tilde{\bm X} =\max( \mathbf{M}^+ \exp (\bm m ) , \epsilon)$.
%
%
An all-pole envelope in frame $k$ is obtained by computing the autocorrelation sequence from $\tilde{\bm X}_k$ via IFFT, and solving the resulting normal equations for the LP polynomial $\bm a_k$ \cite{Makhoul1975-LP-tutorial-review}.

\subsection{Synthesis filter for parallel inference}
\label{sec:stft_synth_filter}

The LP synthesis filter corresponding to $\bm a_k$ is of an infinite impulse response. However, the impulse response can be truncated without noticeable degradation due to the minimum phase property of LP. The synthesis filter can then be applied frame-wise in STFT domain and the time domain filtered signal is obtained simply by inverse STFT (ISTFT).
%
%
More specifically, let the (complex valued) frequency response of the LP polynomial be 
\begin{equation}
    \bm A_k = \mathrm{FFT}\lbrace \bm a_k \rbrace, 
\end{equation}
where the FFT is zero padded to match the number of frequency bins in the STFT.
The corresponding synthesis filter is the inverted frequency response
\begin{equation}
   \bm H_k= \frac{\exp(-\mathrm{i} \angle \bm A_k)}{\max(|\bm A_k|, \epsilon)} ,
\end{equation}
i.e., the numerator contains the sign-inverted phase and the denominator is the  magnitude floored by a small positive $\epsilon$.
Filtering for the entire excitation signal $\hat{\bm e} = G(\bm z, \bm c)$ is applied by multiplication in the STFT domain
\begin{equation}
 \hat{ \bm x} = \mathrm{ISTFT} \lbrace  \mathrm{STFT}\lbrace \hat{\bm e} \rbrace \odot \bm H \rbrace . 
\end{equation}
We use a cosine window as both the STFT analysis window and ISTFT synthesis window.
As all the operations are differentiable, the gradients with respect to the synthetic speech $\partial \hat{\bm x} / \partial G$ and  $\partial \hat{\bm x} / \partial C$ can be computed similarly to \cite{wang2019-neural-source-filter-model}, and used to update  $G$ and $C$. 

\subsection{Network architectures}
\label{sec:network_architectures}

All networks in this paper use a same basic architecture, namely a dilated convolution residual network with gated activations and skip connections to output. In other words, the architecture is similar to the non-causal feedforward WaveNet \cite{Rethage2018-wavenet-speech-denoising}.
%
%
More specifically, a dilated convolution block receives an input tensor $\bm x_i$ from the previous layer and (optionally) a conditioning $\bm c$. Both tensors are of shape $(B, T, R)$, where $B$ is the batch size, $T$ is the number of time-steps, and $R$ is the number of residual channels. A hidden state $\bm h_i$ is computed from the inputs as
\begin{equation}
    \bm h_i = \tanh(\bm W^f_i * \bm x_i + \bm V^f_i \bm c ) \odot \sigma(\bm W^g_i * \bm x_i + \bm V^g_i \bm c) ,
\end{equation}
where $*$ denotes (dilated) convolution. Layer output $\bm y_i = \bm W^o_i \bm h_i + \bm x_i$ is passed  as input to the next convolution layer and  $\bm h_i$ are connected to a post processing layer via skip connections. 
The post-processing layer concatenates the skip connections channel-wise and applies an affine projection, followed by  $\tanh$ non-linearity before a final affine projection to output. Biases are used throughout, but omitted here for brevity.

In $G$ and $C$, the convolutions are zero padded in order to match the timesteps between the input and output.  No zero padding is used in $D$, and the discriminator gradually reduces its input receptive field length into a single timestep. Furthermore, residual connections are not used in $D$.
Notably, the proposed architectures are non-causal, which is permissible for the intended use with utterance level bi-directional acoustic models.


\subsection{Losses}

The current study adapts the loss functions from our previous work \cite{juvela2019-multiscale-gan-synth}.
However, now we can define the losses directly in speech domain, since the LP synthesis filter and overlap-add are integrated to the computation graph. 
Denote the target signal from  by $\bm x$, and the generator model $G$ output passed through the  synthesis filter by $\hat{\bm x}$. 
Additionally, the output of the conditioning model $\bm c = C(\bm m)$ is made accessible to both $G$ and $D$.  
The goal of the generator is to produce $\hat{\bm x}$ that appears to come from the same distribution as $\bm x \sim p_\mathcal{X}$. The conditioning model $C$ collaborates with $G$: it shares the same training objective and obtains its learning signals through $G$.

We use the Wasserstein GAN \cite{Gulrajani2017-wgan-gp-nips}  for our main GAN loss
\begin{equation}
    \Lcal_\mathrm{GAN} = -\E_{x \sim p_\mathcal{X} } \left[ D(\bm x, \bm c) \right] +  \E_{ \hat{x} \sim p_G} \left[ D( \hat{\bm x} , \bm c ) \right] ,
\end{equation}
which the discriminator aims to minimize, and the generator tries to maximize.
%
To keep the discriminator function sufficiently smooth, we use the gradient penalty proposed in \cite{Gulrajani2017-wgan-gp-nips}
\begin{equation}
    \Lcal_\mathrm{GP} = \E_{x \sim p_\mathcal{X}, \hat{x} \sim p_G } \left[ \left(  \norm{ \nabla_{\tilde{\bm x}} D(\tilde{\bm x}, \bm c )} - 1  \right)^2 \right],
\end{equation}
where $\tilde{\bm x} = \varepsilon \bm x + (1-\varepsilon) \hat{\bm x}$ is sampled randomly along the line segment between $\bm x$ and $\hat{\bm x}$.
Additionally, we regularize the discriminator gradient magnitude for the real data samples to avoid a non-convergent training dynamic described in  \cite{Mescheder2018-which-gan-training-methods-converge}
\begin{equation}
    \Lcal_\mathrm{R1} = \E_{x \sim p_\mathcal{X}} \left[  \norm{\nabla_{\bm x} D(\bm x, \bm c )}^2 \right] .
\end{equation}
Finally, we use the following loss term based on the mean squared error of the STFT magnitudes
\begin{equation}
    \Lcal_\mathrm{STFT} = \E \left[ \left( \lvert \mathrm{STFT}\lbrace \bm x \rbrace \rvert -  \lvert \mathrm{STFT} \lbrace \hat{\bm x} \rbrace \rvert \right)^2 \right].
\end{equation}
Notably, the STFT loss uses only the Fourier magnitude, which is insensitive to phase alignment within the short-time frame. All learning signals related to the phase information originate from the discriminator, which in turn does not require parallel data during training. 
The total training objective for $G$ and $C$ is to minimize 
\begin{equation}
    \Lcal_{G,C} =  \lambda_1 \Lcal_\mathrm{STFT} -  \Lcal_\mathrm{GAN},
\end{equation}
 while $D$ attempts to minimize
\begin{equation}
     \Lcal_D = \Lcal_\mathrm{GAN} + \lambda_2 \Lcal_\mathrm{GP} +  \lambda_3 \Lcal_\mathrm{R1}.
\end{equation} 


\section{Experiments}

\subsection{Data}

The experiments were conducted on the ``Nancy'' dataset (from Blizzard Challenge 2011 \cite{King2011-blizzard11}), spoken by a professional female US English voice talent.  The dataset comprises approximately 12000 utterances in normal read style with medium to high expressiveness, totaling to 16\,h\,45\,min of speech. The audio was resampled to 16\,kHz and utterances longer than 8.75 seconds were excluded.  
The remaining 11643 utterances were randomly split to training (11000 utts), validation (200 utts), and test (443 utts) sets. 


\subsection{Training the neural vocoder}

During training, the data was split into one-second segments, such that one training iteration corresponds to processing one second of speech. 
The model was pre-trained in the residual excitation domain, i.e., the inverse filter was applied to the target speech signal to obtain a target excitation. 
Pre-training the model speeds up the training considerably; harmonics started to emerge already at around 60k iterations. 
After excitation model training for 200k iterations, the optimization criteria were switched to the speech signal domain and trained until a total of 1M iterations. This corresponds to approximately 15 full epochs through the dataset.
For discriminator input, we randomly cropped 32 receptive field length segments from the reference and generated signals at each iteration.
The models were optimized using alternating minibatch updates with the Adam optimizer  \cite{Kingma2014-adam} at the default hyperparameters  (LR=1e-4, $\beta_1$=0.9, $\beta_2$=0.999).
Loss weights were chosen experimentally, so that the weighted STFT and GAN losses have roughly the same order of magnitude.  We use $\lambda_1 = 10$ for the STFT loss and    $\lambda_2 = 10$ and $\lambda_3 = 1$ for the gradient penalties.
The network configurations used for the experiments are listed in Table~\ref{tab:network_configs}.
%
%



\begin{table}[th]
  \caption{GELP network configurations. }
  \label{tab:network_configs}
  \centering
  \begin{tabular}{ l | c c c}
     & $G$ & $D$ & $C$  \\
    \midrule
     Residual channels & 64 &  64 & 64  \\
     Skip channels & 64 & 64 & 64 \\
     Filter width & 5 & 5 & 5 \\
     Dilated stacks & 3 & 3 & 2 \\
     Dilation cycle & 8 & 7 & 4 \\
     Residual connection & yes & no & yes \\
     Operation rate & Audio & Audio & Frame \\
    \bottomrule
  \end{tabular}
\end{table}

\subsection{Tacotron synthesis system}
Our Tacotron system is similar to the Tacotron-1 architecture \cite{Wang2017-tacotron}, with minor modifications.
The main modification was to predict mel-spectrograms instead of linear spectrograms as the final output of the system.
The input linguistic features were mono-phonemes extracted using the Combilex lexicon \cite{richmond2009combilex} and represented as one-hot vectors.
Mel-spectrograms were normalized to lie between 0 and 1 using min-max normalization.
The minibatch size was set to 32 and 2 acoustic frames were predicted for each output-step.
The initial learning rate was set to 0.002 and adjusted during the training using the Noam scheme \cite{vaswani2017attention} with 4000 warmup steps.
All the parameters of the systems were optimized using the Adam optimizer.
The system was trained on a single NVIDIA Titan X GPU for 150k steps.
The acoustic features in the Tacotron system use a 80\,Hz frame rate and a 50\,ms window. For the GELP vocoder, the features were upsampled to 200\,Hz rate.

\subsection{Reference methods}

As a simple baseline system for mel-spectrogram-to-waveform synthesis, we first convert from the mel scale to the linear scale via the filterbank pseudo-inverse, and then apply the Griffin-Lim phase recovery algorithm \cite{griffin1984-signal-estimation-from-modified}. This approach avoids the use of a linear-scale spectrogram predictor post-net proposed in \cite{Wang2017-tacotron}, at the cost of some quality loss. However, the intended use of phase recovery in Tacotron-based synthesis is generally not high-quality synthesis, but rather a development tool to quickly check whether the generated acoustic features are meaningful.

 
For a reference autoregressive WaveNet vocoder, we trained a speaker dependent model conditioned on mel-spectrograms, following the ``Wave-30'' configuration from \cite{juvela2019-glotnet-taslp}. The system uses a dilation cycle of 10 (e.g., 1, 2, ..., 512) repeated three times, 64 residual channels, and 256 post-net channels. The model was trained on softmax cross entropy using 8-bit amplitude quantization and $\mu$-law  companding. The conditioning mel-spectrogram features were stacked for a five-frame time context.



\subsection{Listening test}

\begin{figure}
    \centering
    \subfloat{\includegraphics[height=0.75\linewidth, trim={0.0cm, 0.0cm, 0.0cm, 0.0cm}, clip]{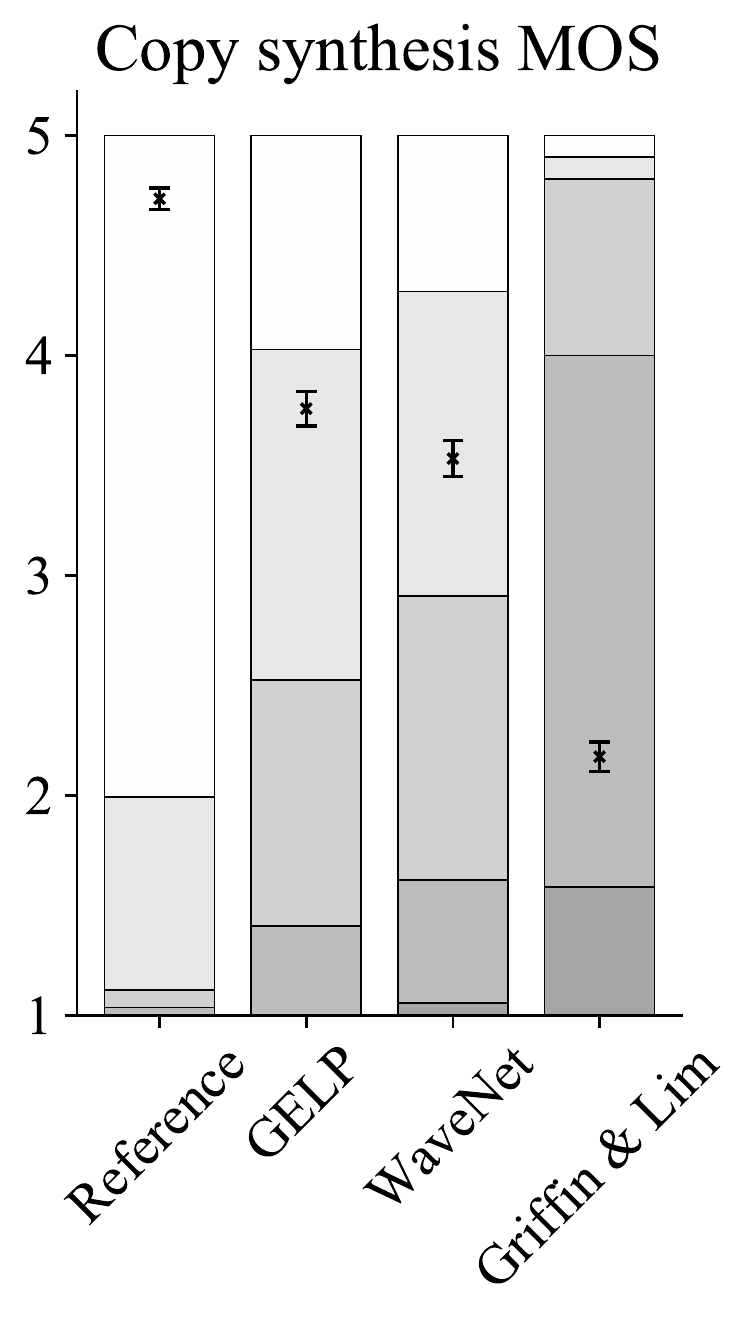}}%
\hfil
\subfloat{\includegraphics[height=0.75\linewidth, trim={0.0cm, 0.0cm, 0.0cm, 0.0cm}, clip]{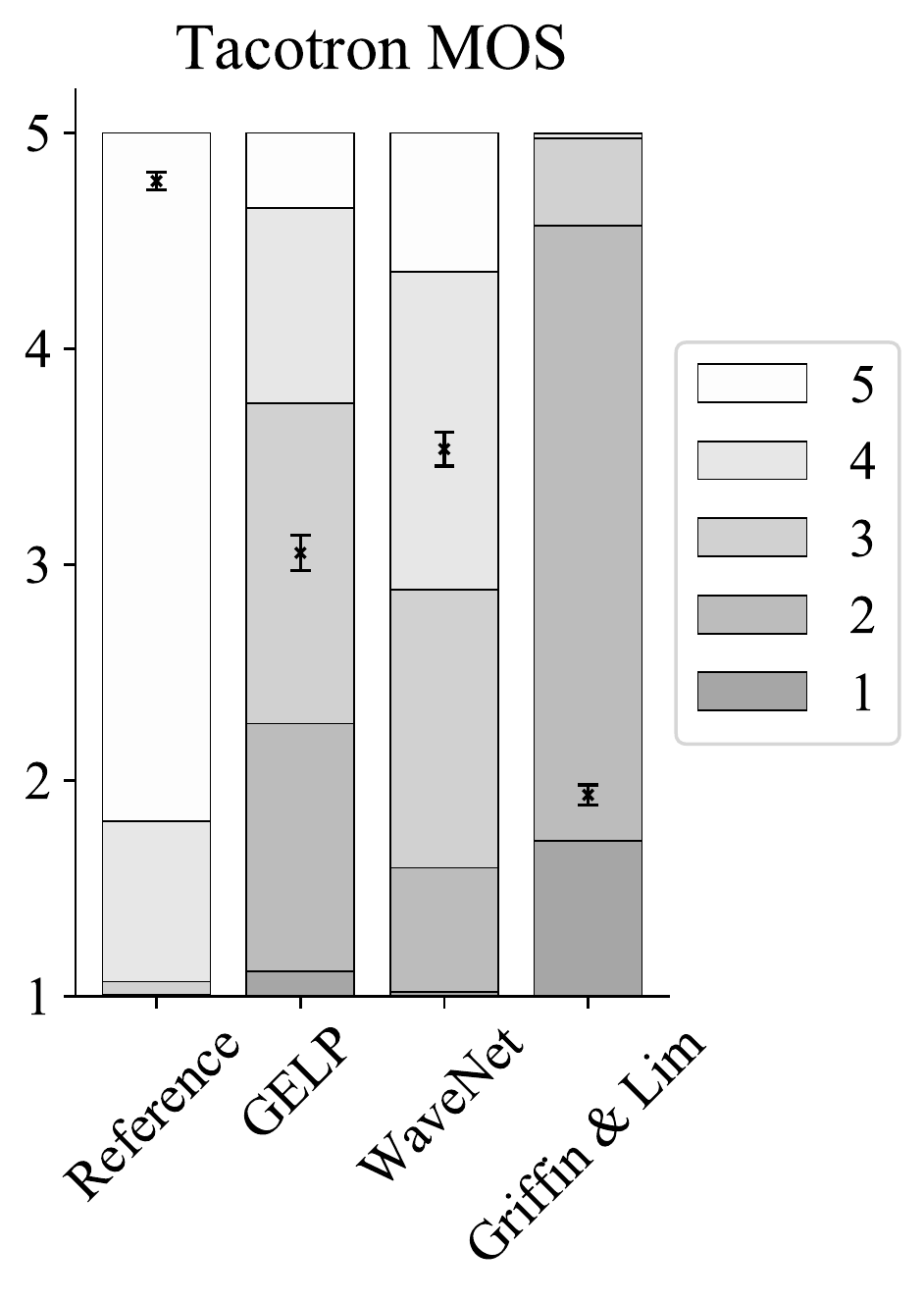}}%
    \caption{Mean opinion scores with 95\% confidence intervals. Stacked score distribution histograms are shown on the background. }
    \label{fig:mos_test_results}
\end{figure}

Subjective quality of the systems was evaluated on the 5 level absolute category rating scale, ranging from 1 (``Bad'') to 5 (``Excellent''). Figure~\ref{fig:mos_test_results} shows the test results as mean opinion scores (MOS) \cite{Itu1996} with the 95\% confidence intervals, along with stacked score distribution histograms. 
 The Mann-Whitney U-test found all differences between systems statistically significant at $p <0.05$ corrected for multiple comparisons. 
The listening tests were split to two groups: copy-synthesis using natural acoustic features, and Tacotron TTS using generated acoustic features.
Both tests include 50 randomly chosen test set utterances, and 20 evaluations were collected for each utterance and system combination. The tests were conducted on the Figure Eight crowd-sourcing platform \cite{figure-eight}.
Demonstration samples are available at
\url{https://ljuvela.github.io/GELP/demopage/}.

\section{Discussion}

In copy-synthesis, GELP achieved higher MOS scores than the reference WaveNet vocoder. However, in the Tacotron TTS experiment, the mean quality of the proposed system dropped below that of WaveNet. 
Quality of the WaveNet-based TTS system is robust to mismatch between natural and generated acoustic features due to the WaveNet's ability to correct its behavior based on previous predictions \cite{valin2019-LPCNet}. 
In contrast, the parallel inference in the proposed GELP model generates the full waveform in a single forward pass, which does not allow corrective, AR-type of feedback. 

On the other hand, a clear benefit of the proposed system is its parallel inference which makes the system significantly faster compared to the autoregressive WaveNet. When synthesizing test set utterances, GELP generated on average 389k samples/second (approximately 24 times real time equivalent rate), on a Titan X Pascal GPU. In comparison, the reference WaveNet with sequential inference generated 217 samples/second on average, which is slower by a factor of 1800.  Although sequential inference can be sped up by careful DSP implementation, it remains  unable to fully leverage the parallel processing power provided by GPUs. Similar speedups over sequential inference have been reported for other parallel neural vocoders \cite{oord2017-parallel-wavenet, ping2018-clarinet-parallel-wave-generation, wang2019-neural-source-filter-model}, but comparison with these is left as future work due to limited available resources for implementation and computation. 


The MOS score measured for the WaveNet vocoder in the current study may seem low compared to the score (about 4.5) reported in \cite{Shen2018-tacotron2}. However, similar studies reporting variations in the WaveNet performance (depending on speaker and acoustic features) are found in the literature: MOS scores around 3.5 have been reported when using mel-filterbank \cite{adiga2018-wavenet-vocoder} or mel-cepstrum \cite{Tamamori2017-wavenet-vocoder} acoustic features. 
Meanwhile, our version of WaveNet has previously achieved MOS scores above 4 using glottal vocoder acoustic features \cite{juvela2019-glotnet-taslp}.
The performance of a parallel waveform generator is likely to increase with the use of a separate pitch predictor model \cite{wang2019-neural-source-filter-model}. However, in this work we chose to limit the acoustic features to mel-spectrogram for their relative simplicity.


\section{Conclusions}

This paper proposed a ``GAN-exicted linear prediction'' (GELP) neural vocoder for fast synthesis of speech waveforms from mel-spectrogram features. The proposed model leverages the spectral envelope information contained in the mel-spectra, and implements an all-pole synthesis filter as part of the computation graph. The model is optimized using STFT magnitude regression and GAN-based losses in time domain. The proposed model has a considerably faster inference speed than a sequential WaveNet vocoder, while outperforming the WaveNet in copy-synthesis quality. 
Future work includes narrowing the quality gap between natural and generated acoustic features.

\section{Acknowledgements}
This study was supported by the Academy of Finland (project 312490), JST CREST Grant Number JPMJCR18A6 (VoicePersonae project), Japan, and by MEXT KAKENHI Grant Numbers (16H06302, 17H04687, 18H04120, 18H04112, 18KT0051), Japan.
We acknowledge the computational resources provided by the Aalto Science-IT project.

\bibliographystyle{IEEEtran}

\bibliography{refs}

\begin{thebibliography}{10}
\providecommand{\url}[1]{#1}
\csname url@samestyle\endcsname
\providecommand{\newblock}{\relax}
\providecommand{\bibinfo}[2]{#2}
\providecommand{\BIBentrySTDinterwordspacing}{\spaceskip=0pt\relax}
\providecommand{\BIBentryALTinterwordstretchfactor}{4}
\providecommand{\BIBentryALTinterwordspacing}{\spaceskip=\fontdimen2\font plus
\BIBentryALTinterwordstretchfactor\fontdimen3\font minus
  \fontdimen4\font\relax}
\providecommand{\BIBforeignlanguage}[2]{{%
\expandafter\ifx\csname l@#1\endcsname\relax
\typeout{** WARNING: IEEEtran.bst: No hyphenation pattern has been}%
\typeout{** loaded for the language `#1'. Using the pattern for}%
\typeout{** the default language instead.}%
\else
\language=\csname l@#1\endcsname
\fi
#2}}
\providecommand{\BIBdecl}{\relax}
\BIBdecl

\bibitem{Zen2013-DNN}
H.~Zen, A.~Senior, and M.~Schuster, ``Statistical parametric speech synthesis
  using deep neural networks,'' in \emph{Proc. ICASSP}, May 2013, pp.
  7962--7966.

\bibitem{Zen2009-SPSS}
H.~Zen, K.~Tokuda, and A.~W. Black, ``Statistical parametric speech
  synthesis,'' \emph{Speech Communication}, vol.~51, no.~11, pp. 1039--1064,
  2009.

\bibitem{fan2014tts}
Y.~Fan, Y.~Qian, F.-L. Xie, and F.~K. Soong, ``{TTS} synthesis with
  bidirectional {LSTM} based recurrent neural networks.'' in
  \emph{Interspeech}, 2014, pp. 1964--1968.

\bibitem{Sotelo2017char2wav}
J.~Sotelo, S.~Mehri, K.~Kumar, J.~F. Santos, K.~Kastner, A.~Courville, and
  Y.~Bengio, ``{Char2Waw}: End-to-end speech synthesis,'' in \emph{ICLR,
  workshop track}, 2017, https://openreview.net/pdf?id=B1VWyySKx.

\bibitem{Wang2017-tacotron}
\BIBentryALTinterwordspacing
Y.~Wang, R.~Skerry-Ryan, D.~Stanton, Y.~Wu, R.~J. Weiss, N.~Jaitly, Z.~Yang,
  Y.~Xiao, Z.~Chen, S.~Bengio, Q.~Le, Y.~Agiomyrgiannakis, R.~Clark, and R.~A.
  Saurous, ``Tacotron: Towards end-to-end speech synthesis,'' in \emph{Proc.
  Interspeech}, 2017, pp. 4006--4010. [Online]. Available:
  \url{http://dx.doi.org/10.21437/Interspeech.2017-1452}
\BIBentrySTDinterwordspacing

\bibitem{oord2016-wavenet}
\BIBentryALTinterwordspacing
A.~van~den Oord, S.~Dieleman, H.~Zen, K.~Simonyan, O.~Vinyals, A.~Graves,
  N.~Kalchbrenner, A.~Senior, and K.~Kavukcuoglu, ``{WaveNet}: A generative
  model for raw audio,'' \emph{arXiv pre-print}, 2016. [Online]. Available:
  \url{http://arxiv.org/abs/1609.03499}
\BIBentrySTDinterwordspacing

\bibitem{Tamamori2017-wavenet-vocoder}
A.~Tamamori, T.~Hayashi, K.~Kobayashi, K.~Takeda, and T.~Toda,
  ``Speaker-dependent {WaveNet} vocoder,'' in \emph{Proc.~Interspeech}, 2017,
  pp. 1118--1122.

\bibitem{Shen2018-tacotron2}
J.~Shen, R.~Pang, R.~Weiss, M.~Schuster, N.~Jaitly, Z.~Yang, Z.~Chen, Y.~Zhang,
  Y.~Wang, R.~Skerry-Ryan, R.~Saurous, Y.~Agiomyrgiannakis, and Y.~Wu,
  ``Natural {TTS} synthesis by conditioning {WaveNet} on mel spectrogram
  predictions,'' in \emph{Proc.~ICASSP}, 2018, pp. 4779--4783.

\bibitem{Wang2018-comparison-of-waveform-generation}
X.~Wang, J.~Lorenzo-Trueba, S.~Takaki, L.~Juvela, and J.~Yamagishi, ``A
  comparison of recent waveform generation and acoustic modeling methods for
  neural-network-based speech synthesis,'' in \emph{Proc.~ICASSP}, 2018, pp.
  4804--4808.

\bibitem{Ramachandran2017-fast-generation-cnn}
P.~Ramachandran, T.~L. Paine, P.~Khorrami, M.~Babaeizadeh, S.~Chang, Y.~Zhang,
  M.~A. Hasegawa-Johnson, R.~H. Campbell, and T.~S. Huang, ``Fast generation
  for convolutional autoregressive models,'' in \emph{Proc.~ICLR (Workshop
  track)}, 2017.

\bibitem{kalbrenner2018-wave-rnn}
\BIBentryALTinterwordspacing
N.~Kalchbrenner, E.~Elsen, K.~Simonyan, S.~Noury, N.~Casagrande, E.~Lockhart,
  F.~Stimberg, A.~van~den Oord, S.~Dieleman, and K.~Kavukcuoglu, ``Efficient
  neural audio synthesis,'' \emph{arXiv pre-print}, 2018. [Online]. Available:
  \url{http://arxiv.org/abs/1802.08435}
\BIBentrySTDinterwordspacing

\bibitem{oord2017-parallel-wavenet}
\BIBentryALTinterwordspacing
A.~van~den Oord, Y.~Li, I.~Babuschkin, K.~Simonyan, O.~Vinyals, K.~Kavukcuoglu,
  G.~van~den Driessche, E.~Lockhart, L.~C. Cobo, F.~Stimberg, N.~Casagrande,
  D.~Grewe, S.~Noury, S.~Dieleman, E.~Elsen, N.~Kalchbrenner, H.~Zen,
  A.~Graves, H.~King, T.~Walters, D.~Belov, and D.~Hassabis, ``Parallel
  {WaveNet}: Fast high-fidelity speech synthesis,'' \emph{arXiv pre-print},
  2017. [Online]. Available: \url{http://arxiv.org/abs/1711.10433}
\BIBentrySTDinterwordspacing

\bibitem{ping2018-clarinet-parallel-wave-generation}
\BIBentryALTinterwordspacing
W.~Ping, K.~Peng, and J.~Chen, ``{ClariNet}: Parallel wave generation in
  end-to-end text-to-speech,'' \emph{arXiv pre-print}, 2018. [Online].
  Available: \url{https://arxiv.org/abs/1807.07281}
\BIBentrySTDinterwordspacing

\bibitem{Prenger2018-waveglow}
\BIBentryALTinterwordspacing
R.~Prenger, R.~Valle, and B.~Catanzaro, ``{WaveGlow}: {A} flow-based generative
  network for speech synthesis,'' \emph{arXiv pre-print}, 2018. [Online].
  Available: \url{http://arxiv.org/abs/1811.00002}
\BIBentrySTDinterwordspacing

\bibitem{wang2019-neural-source-filter-model}
\BIBentryALTinterwordspacing
X.~Wang, S.~Takaki, and J.~Yamagishi, ``Neural source-filter-based waveform
  model for statistical parametric speech synthesis,'' in \emph{Accepted to
  ICASSP}, 2019. [Online]. Available: \url{https://arxiv.org/abs/1810.11946}
\BIBentrySTDinterwordspacing

\bibitem{juvela2018-glotnet-interspeech}
L.~Juvela, V.~Tsiaras, B.~Bollepalli, M.~Airaksinen, J.~Yamagishi, and P.~Alku,
  ``Speaker-independent raw waveform model for glottal excitation,'' in
  \emph{Proc.~Interspeech}, 2018, pp. 2012--2016.

\bibitem{juvela2019-glotnet-taslp}
\BIBentryALTinterwordspacing
L.~Juvela, B.~Bollepalli, V.~Tsiaras, and P.~Alku, ``Glotnet---a raw waveform
  model for the glottal excitation in statistical parametric speech
  synthesis,'' \emph{IEEE/ACM Transactions on Audio, Speech, and Language
  Processing}, 2019. [Online]. Available:
  \url{https://ieeexplore.ieee.org/document/8675543}
\BIBentrySTDinterwordspacing

\bibitem{valin2019-LPCNet}
\BIBentryALTinterwordspacing
J.~S. Jean-Marc~Valin, ``{LPCNet}: Improving neural speech synthesis through
  linear prediction,'' in \emph{Accepted to ICASSP}, 2019. [Online]. Available:
  \url{https://arxiv.org/abs/1810.11846}
\BIBentrySTDinterwordspacing

\bibitem{Hwang2018-LP-wavenet}
\BIBentryALTinterwordspacing
M.-J. Hwang, F.~Soong, F.~Xie, X.~Wang, and H.-G. Kang, ``{LP-WaveNet}: Linear
  prediction-based {WaveNet} speech synthesis,'' \emph{arXiv pre-print}, 2018.
  [Online]. Available: \url{http://arxiv.org/abs/1811.11913}
\BIBentrySTDinterwordspacing

\bibitem{goodfellow2014generative}
I.~Goodfellow, J.~Pouget-Abadie, M.~Mirza, B.~Xu, D.~Warde-Farley, S.~Ozair,
  A.~Courville, and Y.~Bengio, ``Generative adversarial nets,'' in
  \emph{Advances in neural information processing systems}, 2014, pp.
  2672--2680.

\bibitem{juvela2018-synthesis-from-mfcc}
L.~Juvela, B.~Bollepalli, X.~Wang, H.~Kameoka, M.~Airaksinen, J.~Yamagishi, and
  P.~Alku, ``Speech waveform synthesis from {MFCC} sequences with generative
  adversarial networks,'' in \emph{Proc.~ICASSP}, 2018, pp. 5679--5683.

\bibitem{juvela2019-multiscale-gan-synth}
\BIBentryALTinterwordspacing
L.~Juvela, B.~Bollepalli, J.~Yamagishi, and P.~Alku, ``Waveform generation for
  text-to-speech synthesis using pitch-synchronous multi-scale generative
  adversarial networks,'' in \emph{Accepted to ICASSP}, 2019. [Online].
  Available: \url{https://arxiv.org/abs/1810.12598}
\BIBentrySTDinterwordspacing

\bibitem{davis1980mfcc}
S.~Davis and P.~Mermelstein, ``Comparison of parametric representations for
  monosyllabic word recognition in continuously spoken sentences,'' \emph{IEEE
  Transactions on Acoustics, Speech, and Signal Processing}, vol.~28, no.~4,
  pp. 357--366, 1980.

\bibitem{boucheron2012low-bitrate-mfcc-codec}
L.~E. Boucheron, P.~L. De~Leon, and S.~Sandoval, ``Low bit-rate speech coding
  through quantization of mel-frequency cepstral coefficients,'' \emph{IEEE
  Transactions on Audio, Speech, and Language Processing}, vol.~20, no.~2, pp.
  610--619, 2012.

\bibitem{Makhoul1975-LP-tutorial-review}
J.~Makhoul, ``Linear prediction: A tutorial review,'' \emph{Proceedings of the
  IEEE}, vol.~63, no.~4, pp. 561--580, Apr 1975.

\bibitem{Rethage2018-wavenet-speech-denoising}
D.~Rethage, J.~Pons, and X.~Serra, ``A {WaveNet} for speech denoising,'' in
  \emph{Proc. ICASSP}, 2018, pp. 5069--5073.

\bibitem{Gulrajani2017-wgan-gp-nips}
\BIBentryALTinterwordspacing
I.~Gulrajani, F.~Ahmed, M.~Arjovsky, V.~Dumoulin, and A.~C. Courville,
  ``Improved training of {Wasserstein} {GAN}s,'' in \emph{Advances in Neural
  Information Processing Systems 30}, I.~Guyon, U.~V. Luxburg, S.~Bengio,
  H.~Wallach, R.~Fergus, S.~Vishwanathan, and R.~Garnett, Eds.\hskip 1em plus
  0.5em minus 0.4em\relax Curran Associates, Inc., 2017, pp. 5767--5777.
  [Online]. Available:
  \url{http://papers.nips.cc/paper/7159-improved-training-of-wasserstein-gans.pdf}
\BIBentrySTDinterwordspacing

\bibitem{Mescheder2018-which-gan-training-methods-converge}
L.~Mescheder, A.~Geiger, and S.~Nowozin, ``Which training methods for {GAN}s do
  actually converge?'' in \emph{Proc.~ICML}, vol.~80, 2018, pp. 3481--3490.

\bibitem{King2011-blizzard11}
S.~King and V.~Karaiskos, ``The {B}lizzard {C}hallenge 2011,'' in
  \emph{Blizzard Challenge 2011 Workshop}, Turin, Italy, September 2011.

\bibitem{Kingma2014-adam}
D.~P. Kingma and J.~Ba, ``Adam: {A} method for stochastic optimization,'' in
  \emph{Proc. ICLR}, 2015.

\bibitem{richmond2009combilex}
K.~Richmond, R.~A. Clark, and S.~Fitt, ``Robust {LTS} rules with the {C}ombilex
  speech technology lexicon,'' in \emph{Proc. Interspeech}, Brighton, September
  2009, pp. 1295--1298.

\bibitem{vaswani2017attention}
A.~Vaswani, N.~Shazeer, N.~Parmar, J.~Uszkoreit, L.~Jones, A.~N. Gomez,
  {\L}.~Kaiser, and I.~Polosukhin, ``Attention is all you need,'' in
  \emph{Advances in Neural Information Processing Systems}, 2017, pp.
  5998--6008.

\bibitem{griffin1984-signal-estimation-from-modified}
D.~W. Griffin and J.~S. Lim, ``Signal estimation from modified short-time
  fourier transform,'' \emph{IEEE Transactions on Acoustics, Speech, and Signal
  Processing}, vol.~32, no.~2, pp. 236--243, Apr 1984.

\bibitem{Itu1996}
{International Telecommunication Union}, ``{Methods for Subjective
  Determination of Transmission Quality},'' ITU-T SG12, Geneva, Switzerland,
  Recommendation P.800, Aug. 1996.

\bibitem{figure-eight}
{Figure Eight Inc.}, ``Crowd-sourcing platform,''
  https://www.figure-eight.com/, accessed: 2019-04-04.

\bibitem{adiga2018-wavenet-vocoder}
N.~Adiga, V.~Tsiaras, and Y.~Stylianou, ``On the use of {WaveNet} as a
  statistical vocoder,'' in \emph{Proc.~ICASSP}, 2018, pp. 5674--5678.

\end{thebibliography}

\end{document}